\batchmode
\documentstyle[12pt]{article}
\newcommand{\nc}{\newcommand}
\nc{\renc}{\renewcommand}

\nc{\half}{{\textstyle{1\over2}}}
\nc{\etal}{\mbox{\it et al. }}
\nc{\ie}{{\it i.e.}}
\nc{\eg}{{\it e.g.}}

\renc{\thefootnote}{\arabic{footnote}}
\nc{\capt}[1]{{\bf Figure.} {\small\sl #1}}

\nc{\eqs}[2]{\mbox{Eqs.~(\ref{#1},\,\ref{#2})}}
\nc{\eq}[1]{\mbox{Eq.~(\ref{#1})}}

\nc{\figs}[2]{\mbox{Figs.~(\ref{#1},\,\ref{#2})}}
\nc{\fig}[1]{\mbox{Fig~.(\ref{#1})}}

\nc{\tag}[1]{\label{#1} \marginpar{{\footnotesize #1}}}
\nc{\mtag}[1]{\label{#1} \mbox{\marginpar{{\footnotesize #1}}}}
\renc{\baselinestretch}{1.5}
\newlength{\overeqskip}
\newlength{\undereqskip}
\nc{\be}[1]{\begin{equation} \mbox{$\label{#1}$}}
\nc{\bea}[1]{\begin{eqnarray} \mbox{$\label{#1}$}}
\nc{\Section}[2]{\section{#2}\label{#1}}
\nc{\Bibitem}[1]{\bibitem{#1}}
\nc{\Label}[1]{\label{#1}}

\nc{\eea}{\vspace{\undereqskip}\end{eqnarray}}
\nc{\ee}{\vspace{\undereqskip}\end{equation}}
\nc{\bdm}{\begin{displaymath}}
\nc{\edm}{\end{displaymath}}
\nc{\dpsty}{\displaystyle}
\nc{\bc}{\begin{center}}
\nc{\ec}{\end{center}}
\nc{\ba}{\begin{array}}
\nc{\ea}{\end{array}}
\nc{\bab}{\begin{abstract}}
\nc{\eab}{\end{abstract}}
\nc{\btab}{\begin{tabular}}
\nc{\etab}{\end{tabular}}
\nc{\bit}{\begin{itemize}}
\nc{\eit}{\end{itemize}}
\nc{\ben}{\begin{enumerate}}
\nc{\een}{\end{enumerate}}
\nc{\bfig}{\begin{figure}}
\nc{\efig}{\end{figure}}
\nc{\arreq}{&\!=\!&}
\nc{\arrmi}{&\!-\!&}
\nc{\arrpl}{&\!+\!&}
\nc{\arrap}{&\!\!\!\approx\!\!\!&}
\nc{\non}{\nonumber\\*}
\nc{\align}{\!\!\!\!\!\!\!\!&&}

\def\lsim{\; \raise0.3ex\hbox{$<$\kern-0.75em
      \raise-1.1ex\hbox{$\sim$}}\; }
\def\gsim{\; \raise0.3ex\hbox{$>$\kern-0.75em
      \raise-1.1ex\hbox{$\sim$}}\; }
\nc{\DOT}{\hspace{-0.08in}{\bf .}\hspace{0.1in}}
\nc{\Laada}{\hbox {$\sqcap$ \kern -1em $\sqcup$}}
\nc\loota{{\scriptstyle\sqcap\kern-0.55em\hbox{$\scriptstyle\sqcup$}}}
\nc\Loota{{\sqcap\kern-0.65em\hbox{$\sqcup$}}}
\nc\laada{\Loota}
\nc{\qed}{\hskip 3em \hbox{\BOX} \vskip 2ex}

\nc{\real}{{\rm I \! R}}
\nc{\Z}{{\sf Z \!\!\! Z}}
\nc{\complex}{{\rm C\!\!\! {\sf I}\,\,}}
\def\bigid{\leavevmode\hbox{\small1\kern-3.8pt\normalsize1}}
\def\id{\leavevmode\hbox{\small1\kern-3.3pt\normalsize1}}
\nc{\slask}{\!\!\!/}
\nc{\bis}{{\prime\prime}}
\nc{\pa}{\partial}
\nc{\na}{\nabla}
\nc{\ra}{\rangle}
\nc{\la}{\langle}
\nc{\goto}{\rightarrow}
\nc{\swap}{\leftrightarrow}

\nc{\EE}[1]{ \mbox{$\cdot10^{#1}$} }
\nc{\abs}[1]{\left|#1\right|}
\nc{\at}[2]{\left.#1\right|_{#2}}
\nc{\norm}[1]{\|#1\|}
\nc{\abscut}[2]{\Abs{#1}_{\scriptscriptstyle#2}}
\nc{\vek}[1]{{\rm\bf #1}}
\nc{\integral}[2]{\int\limits_{#1}^{#2}}
\nc{\inv}[1]{\frac{1}{#1}}
\nc{\dd}[2]{{{\partial #1}\over{\partial #2}}}
\nc{\ddd}[2]{{{{\partial}^2 #1}\over{\partial {#2}^2}}}
\nc{\dddd}[3]{{{{\partial}^2 #1}\over
	{\partial #2 \partial #3}}}
\nc{\dder}[2]{{{d #1}\over{d #2}}}
\nc{\ddder}[2]{{{d^2 #1}\over{d {#2}^2}}}
\nc{\dddder}[3]{{d^2 #1}\over
	{d #2 d #3}}
\nc{\dx}[1]{d\,^{#1}x}
\nc{\dy}[1]{d\,^{#1}y}
\nc{\dz}[1]{d\,^{#1}z}
\nc{\dl}[1]{\frac{d\,^{#1}l}{(2\pi)^{#1}}}
\nc{\dk}[1]{\frac{d\,^{#1}k}{(2\pi)^{#1}}}
\nc{\dq}[1]{\frac{d\,^{#1}q}{(2\pi)^{#1}}}

\nc{\cc}{\mbox{$c.c.$ }}
\nc{\hc}{\mbox{$h.c.$ }}
\nc{\cf}{cf.\ }
\nc{\erfc}{{\rm erfc}}
\nc{\Tr}{{\rm Tr\,}}
\nc{\tr}{{\rm tr\,}}
\nc{\pol}{{\rm pol}}
\nc{\sign}{{\rm sign}}
\nc{\bfT}{{\bf T }}

\def\GeV{{\rm\ GeV}}

\nc{\cA}{{\cal A}}
\nc{\cB}{{\cal B}}
\nc{\cD}{{\cal D}}
\nc{\cE}{{\cal E}}
\nc{\cG}{{\cal G}}
\nc{\cH}{{\cal H}}
\nc{\cL}{{\cal L}}
\nc{\cO}{{\cal O}}
\nc{\cT}{{\cal T}}
\nc{\cN}{{\cal N}}
\nc{\rvac}[1]{|{\cal O}#1\rangle}
\nc{\lvac}[1]{\langle{\cal O}#1|}
\nc{\rvacb}[1]{|{\cal O}_\beta #1\rangle}
\nc{\lvacb}[1]{\langle{\cal O}_\beta #1 |}
\nc{\bb}{\bar{\beta}}
\nc{\bt}{\tilde{\beta}}
\nc{\ctH}{\tilde{\cal H}}
\nc{\chH}{\hat{\cal H}}
\nc{\1}{\aa}
\nc{\2}{\"{a}}
\nc{\3}{\"{o}}
\nc{\4}{\AA}
\nc{\5}{\"{A}}
\nc{\6}{\"{O}}
\nc{\al}{\alpha}
\nc{\g}{\gamma}
\nc{\Del}{\Delta}
\nc{\e}{\epsilon}
\nc{\eps}{\epsilon}
\nc{\lam}{\lambda}
\nc{\om}{\omega}
\nc{\Om}{\Omega}
\nc{\ve}{\varepsilon}
\nc{\mn}{{\mu\nu}}
\nc{\k}{\kappa}
\nc{\vp}{\varphi}

\nc{\advp}[3]{{\it  Adv.\ in\ Phys.\ }{{\bf #1} {(#2)} {#3}}}
\nc{\annp}[3]{{\it  Ann.\ Phys.\ (N.Y.)\ }{{\bf #1} {(#2)} {#3}}}
\nc{\apl}[3]{{\it  Appl. Phys. Lett. }{{\bf #1} {(#2)} {#3}}}
\nc{\apj}[3]{{\it  Ap.\ J.\ }{{\bf #1} {(#2)} {#3}}}
\nc{\apjl}[3]{{\it  Ap.\ J.\ Lett.\ }{{\bf #1} {(#2)} {#3}}}
\nc{\app}[3]{{\it Astropart.\ Phys.\ }{{\bf #1} {(#2)} {#3}}}
\nc{\cmp}[3]{{\it  Comm.\ Math.\ Phys.\ }{{ \bf #1} {(#2)} {#3}}}
\nc{\cqg}[3]{{\it  Class.\ Quant.\ Grav.\ }{{\bf #1} {(#2)} {#3}}}
\nc{\epl}[3]{{\it  Europhys.\ Lett.\ }{{\bf #1} {(#2)} {#3}}}
\nc{\ijmp}[3]{{\it Int.\ J.\ Mod.\ Phys.\ }{{\bf #1} {(#2)} {#3}}}
\nc{\ijtp}[3]{{\it Int.\ J.\ Theor.\ Phys.\ }{{\bf #1} {(#2)} {#3}}}
\nc{\jmp}[3]{{\it  J.\ Math.\ Phys.\ }{{ \bf #1} {(#2)} {#3}}}
\nc{\jpa}[3]{{\it  J.\ Phys.\ A\ }{{\bf #1} {(#2)} {#3}}}
\nc{\jpc}[3]{{\it  J.\ Phys.\ C\ }{{\bf #1} {(#2)} {#3}}}
\nc{\jap}[3]{{\it J.\ Appl.\ Phys.\ }{{\bf #1} {(#2)} {#3}}}
\nc{\jpsj}[3]{{\it J.\ Phys.\ Soc.\ Japan\ }{{\bf #1} {(#2)} {#3}}}
\nc{\lmp}[3]{{\it Lett.\ Math.\ Phys.\ }{{\bf #1} {(#2)} {#3}}}
\nc{\mpl}[3]{{\it  Mod.\ Phys.\ Lett.\ }{{\bf #1} {(#2)} {#3}}}
\nc{\ncim}[3]{{\it  Nuov.\ Cim.\ }{{\bf #1} {(#2)} {#3}}}
\nc{\np}[3]{{\it  Nucl.\ Phys.\ }{{\bf #1} {(#2)} {#3}}}
\nc{\npps}[3]{{\it  Nucl.\ Phys.\ Proc.\ Suppl.\ }{{\bf #1} {(#2)} {#3}}}
\nc{\pr}[3]{{\it Phys.\ Rev.\ }{{\bf #1} {(#2)} {#3}}}
\nc{\pra}[3]{{\it  Phys.\ Rev.\ A\ }{{\bf #1} {(#2)} {#3}}}
\nc{\prb}[3]{{\it  Phys.\ Rev.\ B\ }{{{\bf #1} {(#2)} {#3}}}}
\nc{\prc}[3]{{\it  Phys.\ Rev.\ C\ }{{\bf #1} {(#2)} {#3}}}
\nc{\prd}[3]{{\it  Phys.\ Rev.\ D\ }{{\bf #1} {(#2)} {#3}}}
\nc{\prl}[3]{{\it Phys.\ Rev.\ Lett.\ }{{\bf #1} {(#2)} {#3}}}
\nc{\pl}[3]{{\it  Phys.\ Lett.\ }{{\bf #1} {(#2)} {#3}}}
\nc{\prep}[3]{{\it Phys.\ Rep.\ }{{\bf #1} {(#2)} {#3}}}
\nc{\prsl}[3]{{\it Proc.\ R.\ Soc.\ London\ }{{\bf #1} {(#2)} {#3}}}
\nc{\ptp}[3]{{\it  Prog.\ Theor.\ Phys.\ }{{\bf #1} {(#2)} {#3}}}
\nc{\ptps}[3]{{\it  Prog\ Theor.\ Phys.\ suppl.\ }{{\bf #1} {(#2)} {#3}}}
\nc{\physa}[3]{{\it  Physica\ A\ }{{\bf #1} {(#2)} {#3}}}
\nc{\physb}[3]{{\it  Physica\ B\ }{{\bf #1} {(#2)} {#3}}}
\nc{\phys}[3]{{\it Physica\ }{{\bf #1} {(#2)} {#3}}}
\nc{\rmp}[3]{{\it  Rev.\ Mod.\ Phys.\ }{{\bf #1} {(#2)} {#3}}}
\nc{\rpp}[3]{{\it Rep.\ Prog.\ Phys.\ }{{\bf #1} {(#2)} {#3}}}
\nc{\sjnp}[3]{{\it Sov.\ J.\ Nucl.\ Phys.\ }{{\bf #1} {(#2)} {#3}}}
\nc{\spjetp}[3]{{\it Sov.\ Phys.\ JETP\ }{{\bf #1} {(#2)} {#3}}}
\nc{\yf}[3]{{\it Yad.\ Fiz.\ }{{\bf #1} {(#2)} {#3}}}
\nc{\zetp}[3]{{\it Zh.\ Eksp.\ Teor.\ Fiz.\  }{{\bf #1}  {(#2)} {#3}}}
\nc{\zp}[3]{{\it Z.\ Phys.\ }{{\bf #1} {(#2)} {#3}}}
\nc{\ibid}[3]{{\sl ibid.\ }{{\bf #1} {#2} {#3}}}
\nc{\rf}[1]{(\ref{#1})}
\nc{\nn}{\nonumber \\*}
\nc{\bfB}{\bf{B}}
\nc{\bfv}{\bf{v}}
\nc{\bfx}{\bf{x}}
\nc{\bfy}{\bf{y}}
\nc{\vx}{\vec{x}}
\nc{\vy}{\vec{y}}
\nc{\oB}{\overline{B}}
\nc{\oI}{\overline{I}}
\nc{\oR}{\overline{R}}
\nc{\rar}{\rightarrow}
\nc{\ti}{\times}
\nc{\slsh}{\hskip-5pt/}
\nc{\sm}{Standard~Model~}
\nc{\MP}{M_{\rm Pl}}
\nc{\tp}{t_{\rm Pl}}
\nc{\ave}{\bar{E}}

\nc{\eff}{{\rm eff}}
\nc{\kk}{\vek{k}}
\nc{\pp}{{\rm p}}
\nc{\ga}{g_{a\gamma}}
\nc{\vv}{\\}
\nc{\eee}{{\bf E}}
\nc{\bbb}{{\bf B}}
\nc{\qcd}{T_{\rm QCD}}
\nc{\G}{\rm \ G}
\def\vec#1{{\bf #1}}

\def\lae{\;^{<}_{\sim} \;} \def\gae{\; ^{>}_{\sim} \;} 

\def\uude{u^{c}u^{c}d^{c}e^{c}}
\def\ell{e^{c}LL}

\begin{document}
{\title{\vskip-2truecm{\hfill {{\small \\
	\hfill \\
	}}\vskip 1truecm}
{\LARGE Naturally Large Cosmological Neutrino 
Asymmetries in the MSSM}}
{\author{
{\sc  John McDonald$^{1}$}\\
{\sl\small Department of Physics and Astronomy,
University of Glasgow, Glasgow G12 8QQ, SCOTLAND}
}
\maketitle
\begin{abstract}
\noindent

       A large neutrino asymmetry is an interesting possibility for cosmology, 
which can have significant observable consequences for nucleosynthesis and 
the cosmic microwave background. However, although it is a possibility, there is no 
obvious reason to expect the neutrino asymmetry to be observably large. Here we note
that if the baryon asymmetry originates via the Affleck-Dine mechanism along a 
d=4 flat direction of the MSSM scalar potential and if the lepton asymmetry originates 
via Affleck-Dine leptogenesis along a d=6 direction, corresponding to the lowest dimension directions conserving R-parity, then the ratio $n_{L}/n_{B}$ 
is naturally in the range $10^{8}-10^{9}$. As a result, a potentially observable neutrino
 asymmetry is correlated with a baryon asymmetry of the order of $10^{-10}$.

\end{abstract}
\vfil
\footnoterule
{\small $^1$mcdonald@physics.gla.ac.uk}

\thispagestyle{empty}
\newpage
\setcounter{page}{1}

\section{Introduction}

         It has long been known that there could exist a large cosmological neutrino 
asymmetry ("degeneracy") \cite{lss,bbn}. This has recently become of particular interest
\cite{kr,lp}, due to its effects on the cosmic microwave background (CMB), which
 will be observed in
 detail by the MAP and PLANCK satellites. In addition, a large neutrino asymmetry can 
also affect Big Bang nucleosynthesis (BBN) \cite{bbn} and large scale structure (LSS) formation \cite{lss}.
 Present CMB, 
nucleosynthesis and LSS bounds can already exclude a range of neutrino asymmetries.

              A number of suggestions for the origin of a large asymmetry have been made 
\cite{foot,ccg}.
In the context of SUSY models, the most natural possibility is probably the Affleck-Dine mechanism \cite{ad}.
However, athough it is possible {\it in principle} to account for a large neutrino asymmetry,
there is no particular reason to expect a very large asymmetry (or, indeed, a large 
asymmetry which is nevertheless small enough to be compatible with present observational
 upper bounds). Recently, an interesting model has been proposed by Casas, Cheng and
 Gelmini (CCG) \cite{ccg}, which is based on  
an Affleck-Dine mechanism in an extension of the minimal SUSY Standard Model
 (MSSM) involving right-handed sneutrinos, and which can account for a large lepton
 asymmetry. In this letter we wish to show that there is good reason to expect a 
large (but not too large) neutrino asymmetry from the Affleck-Dine mechanism in the 
context of the MSSM itself. Our main point is that typically a number of scalar fields 
will have expectation values along flat directions of the MSSM scalar potential 
at the end of inflation. If the baryon asymmetry 
and lepton asymmetry originate from the AD mechanism along different 
flat directions, then the ratio of the baryon number to the lepton number 
will be simply determined by the dimension of the non-renormalizable terms 
responsible for lifting the flat directions. For the R-parity conserving models on which we
 concentrate (which both eliminate dangerous renormalizable B and L violating terms from
 the MSSM 
superpotential and also allow for neutralino dark matter \cite{nilles}), 
the dimension of the non-renormalizable terms is even and so the lowest dimension 
flat directions have $d=4$ and $d=6$ \cite{drt}. If the observed 
B asymmetry originates 
along a $d=4$ direction, then the reheating temperature is fixed to be around 
$10^{8} \GeV$. In this case, if the L asymmetry originates along a $d=6$ direction, 
the ratio of the B to L asymmetry will be $10^{8}-10^{9}$. As a result, a B asymmetry of
about $10^{-10}$ will naturally result in an L asymmetry in the range $0.01-0.1$. This
 mechanism requires no unusually large flat direction vacuum expectation values (VEVs); 
it is simply the conventional AD mechanism taking into account the likelihood that more
 than one flat direction scalar field will have an expectation value at the end of inflation.

\section{Lepton Asymmetry and Present Limits}

            A large neutrino asymmetry has a number of effects on cosmology \cite{lss,bbn}. It changes the
 neutrino decoupling temperature, the primordial production of light elements during BBN, 
the time of matter-radiation equality, the contribution of relic neutrinos to the present 
energy density of the Universe, and alters LSS formation and the CMB. The neutrino 
asymmetry is usually characterized by the neutrino degeneracy parameter $\xi_{\nu} = \mu/T_{\nu}$, 
where $\mu$ is the neutrino chemical potential and $T_{\nu}$ is the neutrino temperature. 
$T_{\nu} = y_{\nu}T_{\gamma}$, where $T_{\gamma}$ is the present photon temperature. $y_{\nu} = (4/11)^{1/3}$ in the absence of 
a neutrino asymmetry and is smaller in the presence of a neutrino asymmetry, as 
the neutrinos decouple at 
a higher temperature \cite{bbn,lp}. The neutrino to entropy ratio is related to the 
degeneracy parameter by (for $(\xi_{\nu}/\pi)^2 \ll 1$)
\be{la1} \eta_{L} = \frac{15}{4 \pi^{4} g(T_{\gamma})} y_{\nu}^{3} \left(\pi^{2} \xi_{\nu} 
+ \xi_{\nu}^{3}\right)   ~,\ee 
where $g(T)$ is the number of light degrees of freedom in thermal equilibrium 
($g(T_{\gamma}) = 2$). BBN imposes a constraint on $\xi_{\nu_{e}}$ \cite{bbn}, 
\be{la2}      -0.06 \lae \xi_{\nu_{e}} \lae 1.1  \;\; ; \;\;   (-4 \times 10^{-3} \lae \eta_{L_{e}} 
\lae 9 \times 10^{-2})   ~.\ee
The upper limit assumes a large asymmetry for $\nu_{\mu}$ or $\nu_{\tau}$. In the absence of such an asymmetry, the upper bound becomes $\xi_{\nu_{e}} \lae 0.14$ ($\eta_{L_{e}} \lae 9.6 \times 10^{-3}$) \cite{bbn}. A more recent analysis tightens this 
to $\xi_{\nu_{e}} \lae 0.09$ ($\eta_{L_{e}} \lae  6.2 \times 10^{-3}$) \cite{kohri}.
 LSS imposes the bound, from the requirement of a sufficiently long matter dominated epoch, 
\be{la3} |\xi_{\nu_{\mu}, \nu_{\tau}}|  \lae 6.9 \;\; ; \;\; (|\eta_{L}| \lae 2.8)    ~.\ee

       A danger for any mechanism generating a  large lepton asymmetry is that 
anomalous B+L violating processes acting on the thermalized lepton number could generate
a baryon asymmetry of a similar order of magnitude \cite{ls}. This is suppressed if the lepton 
number is large enough to prevent electroweak symmetry restoration \cite{linde}, such that the 
sphalerons gain a mass much larger than $T$ \cite{ls,senj}. This possibility is also important
as a way to eliminate dangerous topological defects such as
 domain walls or monopoles \cite{senj}. The most recent estimate for the case of the three 
generation MSSM 
is that $SU(2)_{L} \times U(1)_{Y}$ non-restoration occurs if the lepton asymmetry is larger 
than $n_{L}^{c} = 0.72 T^{3}$ at $T_{ew}$ \cite{senj2}. This translates into a lepton to entropy ratio 
\be{la4} \eta_{L}^{c} = \frac{45 n_{L}^{c}}{2 \pi^{2} g(T) T^{3}} = 0.016     ~,\ee 
where we have used $g(T_{ew}) \approx 100$. Noting that $y_{\nu} \lae (4/11)^{1/3}$, 
this imposes a lower bound on $\xi_{\nu_{i}}$ ($i=e,\;\mu,\;\tau$), 
\be{la5}   \Sigma_{i} \xi_{\nu_{i}} \gae 0.23  ~,\ee 
assuming entropy conservation throughout \cite{jdeg}.
We see that with $\xi_{\nu_{\mu, \; \tau}}$ of the same order of magnitude of $\xi_{\nu_{e}}$, this
can be well within the range of $\xi_{\nu_{e}}$ allowed by BBN. This is very
 important, as in general we would expect the $\nu_{e}$ asymmetry to be of the same 
magnitude as that of $\nu_{\mu,\;\tau}$. Therefore compatibility of the BBN constraint with 
the symmetry non-restoration constraint is essential for the existence of a model which 
can naturally generate a large L asymmetry.
(We also note that, independent of the details of the model,
 the lepton asymmetry can only exist in two 
ranges if it is to be compatible with the observed baryon asymmetry, anomalous B+L violating processes and LSS: $\eta_{L} \lae 10^{-10}$ and $ 0.016 \lae \eta_{L} \lae 2.8$.) 
Future observations by the MAP and PLANCK satellites are expected to probe 
the lepton asymmetry down to $\xi_{\nu} \approx 0.5 $ and $\xi_{\nu} \approx 0.25 $
($\eta_{L} \approx 0.035$ and $\eta_{L} \approx 0.017$) respectively \cite{kr}. 

\section{Flat Directions and the Affleck-Dine Mechanism}

         The scalar potential along a flat direction during inflation has the form \cite{jrev}
\be{ad1} V(\Phi) \approx (m^{2} - c H^{2})|\Phi|^{2} 
+ \frac{\lambda^{2}|\Phi|^{2(d-1)}
}{M_{*}^{2(d-3)}} + \left( \frac{A_{\lambda} 
\lambda \Phi^{d}}{d M_{*}^{d-3}} + h.c.\right)    ~,\ee
where $H$ is the expansion rate of the Universe, $m$ is the usual gravity-mediated 
SUSY breaking scalar mass term ($m \approx 100 \GeV$), $c H^{2}$ is the order $H^2$
 correction to the 
scalar mass due to the energy density of the early Universe \cite{h2} (with $c$ positive 
for Affleck-Dine baryogenesis and typically of order one), $A_{\lambda} = A_{\lambda\;o} 
+ a_{\lambda} H$ (where $A_{\lambda \;o}$ is the conventional gravity-mediated 
SUSY breaking term) and the natural scale of the non-renormalizable terms is $M_{*}$,
 where $M_{*} = M_{Pl}/\sqrt{8 \pi}$ is the supergravity (SUGRA) mass scale. The baryon asymmetry forms when the Affleck-Dine scalar begins to oscillate coherently about zero, which happens at $H \approx m$. 

     The field $\Phi$ is a linear combination of squark, slepton and Higgs fields such that the
 F- and D-term contributions to the renormalizable SUSY scalar potential 
vanish. The flat directions are characterized by the lowest dimension scalar operators
 which have non-zero VEV along the flat direction; these also 
correspond to the non-renormalizable superpotential terms responsible for
 lifting the flat directions and supplying the CP violation responsible for 
generating the asymmetry. The possible R-parity conserving $d=4$ and $d=6$ operators
 are given in Table 2 of Ref. \cite{drt}. The magnitude of the asymmetry generated once
 the
 AD field begins to oscillate coherently is then 
\be{a1}  n \approx m \phi^{2} \sin \;\delta_{CP}              ~,\ee
where $\phi/\sqrt{2}$ is the amplitude of the AD field when it begins to coherently oscillate 
at $H \approx m/c^{1/2}$, $m$ is the mass of the AD scalar and $\delta_{CP}$ is the 
CP violating phase responsible for generating the asymmetry. $\delta_{CP}$ can originate
 in one of two ways. If the A-terms have order $H$ corrections, the phase corresponds to
 the phase difference between $a_{\lambda}$ and $A_{\lambda\;o}$. This is expected in F-term inflation models.
 In minimal D-term inflation models there are no order $H$ corrections to the A-terms \cite{kmr}. In
 this case, the phase is essentially the random initial phase of the AD
scalar relative to the A-terms. In both cases, the most natural possibility is that 
$\delta_{CP} \approx 1$.  We will assume this
 throughout. The initial value of the AD scalar field
is given by
\be{a2}  \phi \approx \left( \frac{2^{d-2}}{(d-1)\lambda^{2}} \right)^{1/(2d-4)}
\left(m^{2} M_{*}^{2(d-3)} \right)^{1/(2d-4)}      ~.\ee      
The present charge to entropy ratio is then 
\be{a3} \eta_{Q} = \frac{2 \pi nT_{R}}{H^{2} M_{Pl}^{2}}  \equiv \frac{T_{R}}{2} 
\frac{c \sin \;\delta_{CP}}{\lambda^{2/(d-2)} } 
\frac{m^{(4-d)/(d-2)} 
M_{*}^{ -2/(d-2)}}{(d-1)^{1/(d-2)}}  ~.\ee
The asymmetries for the $d=4$, $d=6$ and $d=8$ directions are then given by 
\be{a3}  \eta_{4} \approx 7 \times 10^{-11} c_{4} \left(\frac{T_{R}}{10^{8} \GeV}
\right)
\left(\frac{1}{3! \lambda_{4}}\right)  \sin \; \delta_{CP \; 4}    ~,\ee
\be{a4}  \eta_{6} \approx 2 \times 10^{-2} c_{6} \left(\frac{T_{R}}{10^{8} \GeV}\right) 
\left(\frac{1}{5! \lambda_{6}}\right)^{1/2}
\left(\frac{100 \GeV}{m_{6}}\right)^{1/2} 
\sin \; \delta_{CP \; 6}    ~,\ee
\be{a5}  \eta_{8} \approx 22 \; c_{8} 
\left(\frac{T_{R}}{10^{8} \GeV}\right) 
\left(\frac{1}{7! \lambda_{8}}\right)^{1/3}
\left(\frac{100 \GeV}{m_{8}}\right)^{2/3} 
\sin \; \delta_{CP \; 8}    ~,\ee
where we have used as a typical value of the non-renormalizable self-coupling of the 
AD field $\lambda_{d} \approx 1/(d-1)!$, such that the strength of the physical $\Phi$ interaction is determined purely by the mass scale $M_{*}$ \cite{kmr}. Thus 
\be{a6} \frac{\eta_{6}}{\eta_{4}} \approx 3.4 \times 10^{8} 
\left(\frac{3! \lambda_{4}}{(5! 
\lambda_{6})^{1/2}}\right) 
\left(\frac{c_{6}}{c_{4}}\right) 
\left( \frac{100 \GeV}{m_{6}} \right)^{1/2} 
\frac{\sin \delta_{CP\;6}}{\sin \delta_{CP\;4}} \equiv 3.4 \times 10^{8} f_{6} 
~,\ee
where $f_{6}$ is typically of the order of 1. 
Similarly, $\eta_{8}/\eta_{4} \approx 3.0 \times 10^{11} f_{8}$.

     From these we see that, firstly, if the 
observed baryon asymmetry ($\eta_{B\;obs} \approx (3-8) \times 10^{-11}$ \cite{sarkar}) comes from 
a $d=4$ direction, then the paramaters must be close to their maximal or upper bound
 values; $T_{R}$ must be close to the thermal gravitino upper bound $\sim 10^{8} \GeV$\begin{footnote}{Recently it has been suggested that non-thermal production of gravitinos at the end of inflation can impose a much tighter upper bound on $T_{R}$, depending on the details of the inflation model \cite{gtr}.}\end{footnote} \cite{sarkar,grav},
 whilst $\delta_{CP\;4}$ must be close to 1. Secondly, assuming the observed asymmetry is due to $d=4$ baryogenesis, the 
asymmetry from a $d=6$ direction is given by 
\be{a7}   \eta_{6} = (1-3) \times 10^{-2} f_{6}      ~.\ee
Thus an asymmetry of 0.01-0.1 is expected in this case. Therefore if the the L asymmetry
 originates 
from a $d=6$ AD mechanism and the B asymmetry from a $d=4$ AD mechanism, 
a large L asymmetry will exist today, which is naturally within the range of $\eta_{L}$ 
permitted by nucleosynthesis. $SU(2)_{L} \times U(1)_{Y}$ symmetry non-restoration is
 also a natural feature, and the expected range of values is potentially
 observable by MAP and
 PLANCK. On the other hand, if the L asymmetry came from a $d=8$ flat direction, the asymmetry would be
 \be{a7a} \eta_{8} = (9-25) f_{8}   ~.\ee
Although this can be compatible with the LSS upper bound, $|\eta_{L}| \lae 2.8$, 
it would not naturally be compatible with the BBN upper bound on the $\nu_{e}$ asymmetry, $\eta_{L_{e}} 
\lae 0.006$. Thus $d = 6$ AD leptogenesis is favoured. 

\section{Affleck-Dine Cosmology Along Multiple Flat Directions} 

      Usually, the AD mechanism is studied for the case of a single flat direction. 
However, it is likely that more than one flat direction scalar will have a negative 
order $H^2$ correction to its mass squared term. As a result, we can expect several 
flat direction fields to be non-zero at the end of inflation and to begin to oscillate 
coherently once $H \lae m$. 

        The directions which can be simultaneously flat are those characterized by operators
 which do not share any field in common. This can be seen by considering the D-term 
contribution to the scalar potential \cite{nilles}, 
\be{d1}   V_{D} = \frac{g_{\alpha}^{2}}{2} | \Phi_{i}T_{\alpha} \Phi_{i}|^{2}   ~,\ee
where $\alpha$ is the gauge group with generators $T^{\alpha}$ and $\Phi_{i}$ are the MSSM scalar fields.
 For flat directions to be independent, the expectation values of the MSSM fields which
 form the flat direction 
must cancel independently in the D-term. This is only possible if they do not have any field
 in common, since otherwise they would have to be varied simultaneously to keep the D-term zero, implying a single AD field with a single
expectation value. Thus so long as the squark and slepton fields have different
gauge indices or are orthogonal in generation space, the corresponding 
flat directions can simultaneously have non-zero values at the end of inflation. Additional
 constraints arise from the requirement of vanishing renormalizable F-term contributions to
 the scalar potential. This requires that no more than one field in each tri-linear term in the
 MSSM superpotential gains an expectation value. 

     As an explicit example, consider the case where the B asymmetry 
comes from the AD mechanism along the $d=4$ $\uude$ and $QQQL$ directions and the 
L asymmetry comes from the $d=6$ $(\ell)^{2}$ and $(d^{c}QL)^{2}$ directions. Suppose that {\it all} 
scalar fields have negative order $H^2$ corrections to their mass terms. Suppose 
also that the 
reheating temperature is $T_{R} \approx 10^{8} \GeV$. Then we cannot allow a 
B violating AD scalar to be non-zero along a flat direction with $d >4$, since it 
would result in a too large B asymmetry. Most combinations of 'orthogonal' flat directions 
of the MSSM scalar potential will have such $d > 4$ flat directions. However, on scales larger than
 the horizon during inflation we can expect domains with different combinations of flat
 directions to exist, so that anthropic selection will imply that we live in a domain with only
 $d=4$ B violating flat directions. To prevent a B violating $d=6$ flat direction, all the
 9$u^{c}$, 9$d^c$, 3$e^c$, 6$L$ and 18$Q$ fields (including colour, $SU(2)_{L}$ and
 generation indices) must be employed in flat directions in such a way that no $d > 4$ AD
 baryogenesis occurs. For example, the D-term allows the following set of 12 flat direction scalars
\be{ad2} 7(u^{c}d^{c}QQ) + \uude + QQQL + 2(\ell)^{2} + (d^{c}QL)^{2}        ~,\ee
where each flat direction is implicitly characterized by a different combination of gauge and
 generation indices. (The $d=6$ terms in brackets should be thought of as 
$d=3$ terms squared, so that they only involve three fields each.) This set of flat 
directions exhausts all the MSSM squark and slepton fields. 

       This shows that it is possible to have a complete set of flat directions with no
 $d \geq 6$ B violating directions. However, not all of these D-flat directions can be
 consistent with F-flatness conditions, so it is necessary to check that a complete set of D-  {\it and}
 F-flat directions with no
 $d \geq 6$ B violating directions exists. An example is given by 
\be{ee1} u^{c\;1}_{2}u^{c\;2}_{3}d^{c\;3}_{3}e^{c}_{3} + 
 (d^{c\;3}_{2}u^{3}_{1}l_{2})^{2}    ~\ee
corresponding to a $d=4$ B violating direction from a gauge-invariant monomial of the
 form\begin{footnote}{The $d=4$ B violating operator responsible for lifting this flat direction need not be additionally suppressed in order to avoid rapid proton decay, as the operator is composed purely of second- and third-generation fields \cite{pdecay}.}\end{footnote} $\uude$ 
and a $d=6$ L violating direction from $(d^{c}QL)^{2}$  (where superscripts
 denote colour indices and subscripts denote flavour indices.)  Choosing a flavour basis with the down quark and lepton Yukawa matrices diagonalized, 
the squark
fields which can form flat directions in addition to those in \eq{ee1} once F-flatness
 conditions are imposed are of the form $u^{c\;i}, \;d^{c\;i}$  ($i=1,2$) and $Q^{3}$. Since
 there is no gauge invariant monomial which can be constructed using these fields, there are no other B violating flat directions. 

         Once the B and L asymmetries are established at $H \approx m$, the
Universe is dominated by coherently oscillating scalar field condensates, corresponding to
the inflaton and the various AD scalars. One difference between the MSSM AD
 mechanism presented here and that of CCG based on right-handed sneutrinos is that 
the AD condensate in our case carries an asymmetry of left-handed sneutrinos and so 
serves as a source term for finite-density $SU(2)_{L} \times U(1)_{Y}$ breaking \cite{linde,jdeg}. 
Thus $SU(2)_{L} \times U(1)_{Y}$ non-restoration occurs throughout. In the CCG model, 
$SU(2)_{L} \times U(1)_{Y}$ breaking occurs only once the right-handed sneutrino 
condensate decays to create a density of left-handed neutrinos. This must occur 
before the inflaton has thermalized, otherwise sphaleron processes would convert the 
lepton asymmetry from the decaying sneutrino field into a large baryon asymmetry. This
 leads to additional constraints on the model \cite{ccg}. In our model 
there are no constraints from inflaton thermalization. In addition, in the CCG model an
 observably large asymmetry can be achieved only if the right handed sneutrino field
 expectation value at the end of inflation is very
 large, $ \tilde{\nu}_{R} \gae 10^{17} \GeV$ \cite{ccg}. This is difficult to achieve, since it
 requires a
 heavy suppression of non-renormalizable operators along the flat direction. 
In our case a large neutrino asymmetry is generated via natural 
expectation values for the AD scalars after inflation.

\section{Conclusions}

       We have discussed the possibility of the AD mechanism being responsible for 
both baryogenesis and leptogenesis, with the B asymmetry originating along  
a $d=4$ direction of the MSSM scalar potential and the L asymmetry 
along a $d=6$ direction. This naturally correlates a lepton asymmetry 
$\eta_{L} \approx 0.01-0.1$ with the observed baryon asymmetry
$\eta_{B} \approx 10^{-10}$. This is compatible with present nucleosynthesis and 
structure formation constraints and is in the range detectable by the MAP and PLANCK 
satellites in the future. It also implies finite density $SU(2)_{L} \times U(1)_{Y}$ 
non-restoration, which is essential for the consistency of the model. The model is
 dependent upon anthropic selection to eliminate dangerous B violating flat directions; however, a domain structure of the Universe, with
 different combinations of flat directions on scales much larger than the present
 horizon, will be a natural feature of the MSSM in the context of inflation models.
 Therefore it is quite natural that we find ourselves in a domain determined by
 anthropic selection.

    The fact that a large lepton asymmetry is a very natural feature of the 
Affleck-Dine mechanism in the MSSM, in which several flat direction 
scalar condensates typically form after inflation, should be a source of 
encouragement to those interested in the cosmological effects of a large neutrino asymmetry.

\subsection*{Acknowledgements}   This research has been supported by the PPARC.

\end{document}